\documentclass[aps,letterpaper,nofootinbib,preprint,showpacs,preprintnumbers,amsmath,amssymb]{revtex4}%
\usepackage{latexsym}%
\usepackage{epsf}%
\usepackage[hypertex]{hyperref}%

\def\cF{{\mathcal F}}

\def\cL{{\mathcal L}}

\def\cM{{\mathcal M}}

\def\atanh{{\text{Arctanh}}}

\newcommand{\beq}{\begin{equation}}
\newcommand{\beqn}{\begin{equation}\nonumber}
\newcommand{\eeq}{\end{equation}}
\newcommand{\bea}{\begin{eqnarray}}
\newcommand{\bean}{\begin{eqnarray}\nonumber}
\newcommand{\eea}{\end{eqnarray}}

\begin{document}

\begin{center}
{\bf{\Large A Lagrangian Description of Thermodynamics}}
\bigskip
\bigskip

{{Cenalo Vaz\footnote{e-mail address: Cenalo.Vaz@UC.Edu}}
\bigskip

{\it RWC and Department of Physics,}\\
{\it University of Cincinnati,}\\
{\it Cincinnati, Ohio 45221-0011, USA}}
\end{center}
\bigskip
\bigskip
\medskip

\centerline{ABSTRACT}
\bigskip\bigskip

The fact that a temperature and an entropy may be associated with horizons in semi-classical 
general relativity has led many to suspect that spacetime has microstructure. If this is 
indeed the case then its description via Riemannian geometry must be regarded as an effective 
theory of the aggregate behavior of some more fundamental degrees of freedom that remain 
unknown, in many ways similar to the treatment of fluid dynamics via the Navier-Stokes equations. 
This led us to ask how a geometric structure may naturally arise in thermodynamics or 
statistical mechanics and what evolution may mean in this context. In this article we argue 
that it is possible to view thermodynamic processes as the evolution of a dynamical system, 
described by a quadratic Lagrangian and a metric on the thermodynamic configuration space. 
The Lagrangian is an invariant distance between equilibrium thermodynamic states as defined 
by the metric, which is straightforwardly obtained from a complete set of equations of state. 
\bigskip\bigskip
\vfill\eject

\section{Introduction}

Research over the past few decades has shown that when the principles of quantum mechanics are 
combined with a geometric theory of gravity in the presence of horizons, spacetime exhibits thermodynamic 
behavior \cite{bch73,haw75}. Without reference to the gravitational field equations, both a 
temperature and an entropy can be associated with any horizon in the following sense: whenever 
a horizon is present there is also a law, similar to the first law of thermodynamics, relating 
changes in the entropy to changes in the energy and other work terms (when appropriate). One 
attempts to understand this peculiar result by noting that horizons act as one way screens, 
causing information from a portion of the spacetime to become inaccessible to an observer on 
one side of it \cite{bek73}. For example, an exterior observer cannot ``see'' what is going on 
within the event horizon of a black hole and a Rindler observer will never have access to the 
portion of Minkowski spacetime on the ``other side'' of her horizon.  Notwithstanding this insight, 
the result itself remains highly non-trivial and intriguing because the appearance of a thermodynamic 
description suggests the existence of spacetime microstructure \cite{pad10}. In other words, it 
is possible that our description of spacetime is actually a description of the aggregate behavior 
of some as yet undiscovered microscopic degrees of freedom. 

This has led some investigators to suggest that gravity and perhaps even a continuum ``spacetime'' 
are emergent phenomena \cite{sak68,jac95,pad06,hor09,ver10}. However, fundamental to the description 
of spacetime, at least in the current approaches, is the metric, which is determined by the 
equations of Einstein's relativity or of some generalization thereof, such as the Lanczos-Lovelock
models in higher dimensions \cite{lov71}. In an emergent picture of spacetime the metric must arise 
as an effective description, that is, out of the statistical mechanics of the more fundamental 
degrees of freedom of which it is made. How does this happen and how are its dynamics as 
encapsulated in, say, Einstein's equations to be reconciled with such a description? This motivates 
us to take a closer look at the structural relationships between our descriptions of dynamical 
systems and of thermodynamic systems. 

In this paper we ask the simpler question of whether there is anything truly ``dynamical'' about 
thermodynamics. That is, we ask if it is possible to describe thermodynamic processes as trajectories 
in the thermodynamic phase space in a way that is completely analogous to the description of a 
dynamical system. We argue that every thermodynamic system can be reformulated as a Hamiltonian 
system, analogous to the one describing a point particle moving in a curved space. The curved 
space is the configuration space of the thermodynamic system and its local geometry is captured 
by a matrix of functions, which incorporate all the physical attributes of the substance. If the 
matrix of functions is invertible, a quadratic Lagrangian description can be obtained. This leads 
to a natural interpretation of the matrix of functions as a metric on the configuration space: 
it determines a ``distance'' between equilibrium thermodynamic states. 

More precisely, the first law of Thermodynamics for quasi-static processes can always be written 
in the form
\beq
df = p_\mu dq^\mu
\label{thermo1}
\eeq
where $f$ represents a thermodynamic potential or the entropy and $q^\mu$ represent $n$ intensive or 
extensive variables of the system (including, possibly, the entropy), while each conjugate generalized 
force, 
\beq
p_\mu = \partial_\mu f,
\eeq 
is extensive if $q^\mu$ is intensive and vice-versa. As a simple example, consider the first 
law for an ideal gas in the form
\beq
du = Tds - pdv
\eeq
where $u$ is the specific internal energy, $s$ the specific entropy, $v$ the specific volume of 
the gas and $T,p$ are its temperature and pressure respectively. In this description the internal 
energy has the privileged role of $f$ in \eqref{thermo1} but an equivalent form
\beq
ds = \frac {du}T + \frac pT dv
\eeq
sees the specific entropy $s$ playing that role. Therefore there is nothing special about $f$ and 
the thermodynamic phase space is naturally odd dimensional. The first law can be viewed as the statement 
that physical thermodynamic processes may only occur on the hypersurface determined by the vanishing of 
the contact form 
\beq
\omega := df - p_\mu dq^\mu.
\label{contact}
\eeq
Rajeev \cite{raj08} has noted that there are many systems of variables in which the contact form will assume 
its canonical expression, as above, and all of them will be determined by requiring that
\beq
\sigma(f,p,q) (df - p_\mu dq^\mu) = df' - P_\mu dQ^\mu
\eeq
for any arbitrary function $\sigma$ on the $2n+1$ dimensional thermodynamic phase space. Transformations 
that take $(f,p, q)$ to $(f',P,Q)$ are the Legendre transformations of thermodynamics. Familiar and useful 
examples of Legendre transformations are transformations to the enthalpy, the Helmholz free energy and the 
Gibbs free energy. 

Here we will consider a restricted set of transformations, namely transformations 
of the $2n$ dimensional subspace $(q^\mu,p_\mu) \rightarrow (Q^\mu,P_\mu)$ that leave the right hand side 
of \eqref{thermo1} form invariant up to the addition of an exact form, {\it i.e.,}
\beq
df = p_\mu dq^\mu = P_\mu dQ^\mu + d \sigma(p,q)
\eeq
These are canonical transformations, under which $f \rightarrow f' = f -\sigma$. In section II we 
will determine a general equation for the generator of these transformations. Because they are 
canonical, the characteristic curves of the generating function will satisfy Hamilton's equations. 
In section III, we confine ourselves to a special subset of these transformations for which an explicit 
form of the generating function can be obtained. This solution is given in terms of a matrix of 
functions. If the matrix is invertible it behaves like a metric on the configuration space and we show 
that there is a Lagrangian description for the evolution. This description is analogous to the Lagrangian 
description of a particle moving in a curved space defined by the metric. We argue that suitable metrics 
may be constructed starting from a complete set of equations of state, which in turn may be 
experimentally determined but are usually derived from a statistical model. Some examples are worked 
out in section IV and we end with a brief discussion of our aims and conclusions in section V.

\section{Canonical Transformations for Thermodynamics}

Let $U$ be a local patch with coordinates $(q^\mu,p_\mu)$ of a $2n$ dimensional manifold $\cM$ on which 
is defined the contact structure \eqref{contact}. Consider a structure preserving transformation of the 
$2n$ coordinates, {\it i.e.,} consider $p_\alpha \rightarrow P_\alpha(q^\mu,p_\mu)$, $q^\alpha \rightarrow 
Q^\alpha(q^\mu,p_\mu)$ such that 
\beq
df = p_\mu dq^\mu  = P_\mu dQ^\mu + d\sigma(q,p)
\eeq
where $\sigma(q,p)$ is an arbitrary function of the coordinates, which gives
\bea
p_\mu &=& P_\alpha \frac{\partial Q^\alpha}{\partial q^\mu} + \frac{\partial \sigma}{\partial q^\mu}\cr\cr
0 &=& P_\alpha \frac{\partial Q^\alpha}{\partial p_\mu} + \frac{\partial \sigma}{\partial p_\mu}
\label{cond}
\eea
Transformations that satisfy these conditions are canonical transformations. An infinitesimal canonical 
transformation can be written as
\beq
p_\mu  \rightarrow P_\mu = p_\mu + \delta \tau~ \eta_\mu,~~ q^\mu \rightarrow Q^\mu = q^\mu + \delta \tau~
\varepsilon^\mu
\label{infinitesimal}
\eeq
where $\eta_\mu(q,p)$ and $\varepsilon^\mu(q,p)$ are functions of $(q,p)$ and $\tau$ is 
some parameter. Let $\sigma(q,p) = \delta \tau~\lambda(q,p)$, then applying the conditions 
\eqref{cond} it is easy to see that 
\bea
&&\eta_\mu = - p_\alpha \frac{\partial \varepsilon^\alpha}{\partial q^\mu} - \frac{\partial \lambda}{\partial 
q^\mu}\cr\cr
&&p_\alpha \frac{\partial \varepsilon^\alpha}{\partial p_\mu} + \frac{\partial \lambda}{\partial p_\mu} = 0
\eea
The solution of these equations can be determined in terms of the function $F = p_\alpha \varepsilon^\alpha + 
\lambda$ as
\beq
\eta_\mu = -\frac{\partial F}{\partial q^\mu},~~ \varepsilon^\mu = \frac{\partial F}{\partial p_\mu}
\eeq
so that $F$ is the generating function of the infinitesimal canonical transformations. Finite transformations 
may be recovered by composing such infinitesimal transformations, that is by determining the integral curves 
of the vector field 
\beq
V = \eta_\mu \frac{\partial}{\partial p_\mu} + \varepsilon^\mu \frac{\partial}{\partial q^\mu}
\eeq
Thus the generating function $F$ defines a one parameter family of curves. These are its characteristic 
curves and, in terms of the mock ``time'' parameter, $\tau$, introduced in  \eqref{infinitesimal}, they 
satisfy the ordinary differential equations
\beq
\dot q^\mu = \frac{\partial F}{\partial p_\mu},~~ \dot p_\mu = -\frac{\partial F}{\partial q^\mu},
\label{motion}
\eeq
where the over dot represents a derivative with respect to $\tau$. We have thus recovered precisely 
Hamilton's equations giving the classical trajectories of a particle described by the ``Hamiltonian'' 
$F$,
\beq
F = p_\mu \dot q^\mu + \lambda = p_\mu \frac{\partial F}{\partial p_\mu} + \lambda.
\label{Feqn}
\eeq
If $\lambda=0$ then $f$ does not change during the process and $F$ is extensive in the ``momenta'', 
\beq
F(q^\mu,\mu p_\mu) = \mu F(q^\mu,p_\mu),
\eeq 
but if $\lambda \neq 0$, then 
\beq
f \rightarrow f' = f - \delta \tau~ \lambda
\eeq
so that the transformations induce a change in $f$ according to 
\beq
\dot f = - \lambda.
\label{fdot}
\eeq
Now the physical hypersurface of $\cM$ is the one on which the contact form vanishes, {\it i.e.,}
\beq
\dot f = p_\mu \dot q^\mu = p_\mu \frac{\partial F}{\partial p_\mu} = - \lambda + F
\eeq
by \eqref{Feqn} and so it is determined by the condition that $F=0$. 

\section{A Restricted Class of Transformations}

Let us henceforth consider transformations for which $\lambda = \lambda(q)$. In that case, the 
following is a formal solution of \eqref{Feqn} for $F$,
\beq
F = \sqrt{g^{\mu\nu}(q) p_\mu p_\nu} + \lambda(q),
\eeq
where $g^{\mu\nu}(q)$ is an $n \times n$ dimensional matrix of arbitrary functions on the configuration 
space. The constraint, $F=0$, defining the physical hypersurface translates into
\beq
\sqrt{g^{\mu\nu}(q) p_\mu p_\nu} = -\lambda(q)
\label{const}
\eeq
and the evolution equations for this system are found from \eqref{motion} to be
\bea
&&\dot q^\mu = \frac{\partial F}{\partial p_\mu} = \frac{g^{\mu\nu} p_\nu}{\sqrt{g^{\mu\nu}p_\mu p_\nu}}
= - \lambda^{-1}(q)g^{\mu\nu} p_\nu \cr\cr
&&\dot p_\mu = -\frac{\partial F}{\partial q^\mu} = -\frac 12 \lambda^{-1}(q) {g^{\alpha\beta}}_{,\mu} 
p_\alpha p_\beta- \partial_\mu \lambda
\label{eqmot}
\eea
where use has been made of \eqref{const}. If in addition the matrix of functions $g^{\mu\nu}(q)$ is 
invertible, then it may be thought of as a metric on the configuration space of the thermodynamic system, 
for one finds that the constraint in \eqref{const} turns into the condition
\beq
g_{\mu\nu} \dot q^\mu \dot q^\nu = 1
\eeq
and this gives a natural distance 
\beq
d\tau^2 = g_{\mu\nu} dq^\mu dq^\nu
\eeq
between equilibrium states of the system. Furthermore, because $\tau$ is an arbitrary parameter, 
the right hand side of the above equation is invariant under coordinate transformations implying 
that the metric $g_{\mu\nu}$ transforms as a covariant tensor of rank two. Using the first of 
\eqref{eqmot}, one has
\beq
p_\mu = -\lambda(q) g_{\mu\nu}(q) \dot q^\nu
\eeq
and the second equation simplifies to
\beq
\frac{dp_\mu}{d\tau} - \lambda^{-1}g^{\kappa\nu}\Gamma^\lambda_{\mu\nu}p_\lambda p_\kappa 
= - \partial_\mu \lambda
\eeq
where ``$\Gamma^\lambda_{\mu\nu}$'' is the Christoffel connection defining parallel transport induced by 
the metric $\widehat g$. These equations can be recovered by extremizing 
the reparametrization invariant action
\beq
S = \int d\tau \cL(q(\tau),\dot q(\tau),\tau) = -\int d\tau \lambda(q) \sqrt{g_{\mu\nu}(q) \dot 
q^\mu(\tau) \dot q^\nu(\tau)}
\label{action}
\eeq
where the momentum is defined in the usual way as
\beq
p_\mu = \frac{\partial \cL}{\partial \dot q^\mu}.
\eeq
The action \eqref{action} determines the minimum length path between the initial and final 
states in the thermodynamic configuration space with the conformal metric 
\beq
\widetilde g_{\mu\nu} = \lambda^2(q) g_{\mu\nu}.
\eeq
The mock time parameter, $\tau$, can be related to $f$ by choosing $\lambda$. Taking 
$\lambda = \lambda_0$ (constant) forces $\tau$ to be proportional to $f$ according to 
\eqref{fdot}. In this gauge, the action describing the evolution of our thermodynamic system 
becomes
\beq
S = - \lambda_0 \int \sqrt{g_{\mu\nu}(q)~ dq^\mu~ dq^\nu} = -\lambda_0 \int d\tau,
\label{finalaction}
\eeq
which is precisely the action governing the evolution of a particle of mass $\lambda_0$ in 
a curved space defined by the metric $\widehat g$. 

The metric necessarily contains all the information about the substance, so it must be 
determined from appropriate additional considerations. Once determined, all processes are described 
by its geodesics. One possible choice of metric, that we will not adhere to in the examples of the next 
section, is obtained by noting that the quadratic form in \eqref{finalaction} must be required to be 
positive definite. This is guaranteed by the second law of thermodynamics, {\it i.e.,} by the concavity 
of the entropy function, $s$, if we take 
\beq
g^R_{\mu\nu} = -\frac{\partial^2 s}{\partial q^\mu \partial q^\nu}.
\eeq
This metric has been proposed by Ruppeiner\cite{rup95} as it arises naturally out of classical 
thermodynamic fluctuation theory. However, it requires a knowledge of the entropy as a function of 
the extensive variables, which is tantamount to a complete knowledge of the thermodynamics. Likewise,
for the Weinhold metric \cite{wei76}, which requires a knowledge of the internal energy.

From a thermodynamic point of view, a substance is characterized by a set of equations of state 
that are experimentally determined. For example, an ideal gas is completely characterized by the 
law of Boyle, Charles and Gay-Lussac together with a relationship between its internal energy and 
its temperature. These relations, although historically experimental, are also obtained 
directly from statistical models. In the statistical approach, a theoretical model is constructed and a 
partition function, $Z(\beta,q^i)$, where $\beta$ is the inverse temperature, is obtained, from 
which the internal energy and conjugate forces are determined according to
\beq
u = - \frac{\partial \ln Z}{\partial\beta},~~ \cF_i(\beta,q) = \frac 1\beta \frac{\partial \ln Z}
{\partial q^i}.
\eeq
These $n$ equations of state suffice to construct a family of invertible metrics all of which define a 
quadratic form of the kind $g^{\mu\nu} p_\mu p_\nu$, whose value is a constant on the physical hypersurface. 
More precisely, suppose that we have $n$ independent equations of state in the form
\beq
f_n^\mu (q) p_\mu = p_n
\eeq
where $p_n$ are constant. The matrix $\widehat f$ must be invertible so that it is possible to recover 
the momenta from the equations of state, 
\beq
p_\mu = (f^{-1})^n_\mu (q)p_n.
\eeq 
If we take the $p_n$ to define 
an orthogonal basis in a real vector bundle over the configuration space, with an invertible, constant 
matrix $\eta_{mn}$, then $f^\mu_n$ can be thought of as a vielbein, and one has the natural metric
\beq
g^{\mu\nu} = \eta^{mn} f_m^\mu f_n^\nu
\label{geng}
\eeq
on $\cM$. Furthermore, on the physical hypersurface, {\it i.e.,} when $F=0$,
\beq
\lambda = -\sqrt{g^{\mu\nu} p_\mu p_\nu} = -\sqrt{\eta^{mn} p_m p_n} =\lambda_0
\eeq
is a constant determined by $\widehat \eta$ and $p_n$ (we shall henceforth drop the subscript ``0''). 
Taking $\widehat \eta$ to be diagonal, we find
that $g^{\mu\nu}$ is actually an $n$ parameter family of metrics. As we will see in the following 
examples, they serve to parameterize the solutions.

We assume that a complete characterization of the substance by means of $n$ independent equations of 
state of the above form is available. Using these equations as our starting point, we assemble a metric 
according to \eqref{geng}, which automatically gives a constant $\lambda$ on the physical hypersurface. 
Below we illustrate the formalism for some common systems.

\section{Examples}

Because the experimentally determined equations of state do not generally include the entropy, 
we single it out by letting $f=s$ in the following examples. This is not necessary, but it is 
convenient for our purposes. Further, all of our examples are two dimensional and we will 
have two independent equations of state, so we take 
\beq
\eta_{mn} = \left(\begin{matrix}
\alpha^2 & 0 \cr
0 & \beta^2
\end{matrix}\right)
\eeq
for arbitrary $\alpha$ and $\beta$.

\subsection{The ideal gas}

Take $f=s$, the specific entropy of the gas, and the first law in the form 
\beq
ds = \frac 1T du + \frac pT dv
\eeq
In this representation, the coordinates are $(u,v)$, the momenta $p_u=1/T$, $p_v=p/T$ 
and the ideal gas can be characterized by the two equations of state
\bea
p_u u &=& \frac{gk}2\cr\cr
p_v v &=& k
\label{stateig}
\eea
where $k$ is Boltzmann's constant and $g$ is the number of degrees of freedom per molecule. 
According to \eqref{geng} this gives the two parameter family of metrics
\beq
g^{\mu\nu} = \left(\begin{matrix}
\frac{u^2}{\alpha^2} & 0\cr
0 & \frac{v^2}{\beta^2}
\end{matrix}\right),
\label{gig}
\eeq
and, from the on-shell condition,
\beq
\lambda = - \sqrt{\frac{g^2 k^2}{4\alpha^2} + \frac{k^2}{\beta^2}}
\label{lig}
\eeq
We find the evolution equations
\bea
\dot u = -\frac{p_u u^2}{\alpha^2\lambda},~~ \dot v = -\frac{p_v v^2}{\beta^2\lambda},\cr
\dot p_u = \frac{p_u^2 u}{\alpha^2\lambda},~~ \dot p_v = \frac{p_v^2 v}{\beta^2\lambda},
\eea
which will be seen to directly reproduce the equations of state and, furthermore, using the fact 
that $s=-\lambda\tau$, give the solutions
\beq
u = u_0 \exp\left[\frac{gks}{2\alpha^2\lambda^2}\right],~~ v = v_0 \exp\left[\frac{ks}{\beta^2\lambda^2}
\right],
\label{solsig}
\eeq
together with corresponding solutions for $p_u$ and $p_v$, which follow from \eqref{stateig}. 
They can be inverted and the constants $\lambda$, $\alpha$ and $\beta$ eliminated using 
\eqref{lig} to give 
\beq
s=k \ln \left[\left(\frac{u}{u_0}\right)^{g/2} \frac v{v_0}\right].
\eeq
The solutions in \eqref{solsig} describe a two parameter family of ideal gas processes. For instance 
the choice $\alpha^2 = gk/2$, $\beta^2 = k$ makes \eqref{gig} equal to the Ruppeiner (entropy) metric 
and describes a constant pressure process. A constant temperature process can be recovered in the 
limit as $\alpha \rightarrow \infty$ and a constant volume process in the limit as $\beta\rightarrow 
\infty$. It is worth noting that these trajectories are also recovered as geodesics of the metric 
in \eqref{gig}, or from the action
\beq
S = -\lambda \int d\tau \sqrt{\dot x^2 + \dot y^2},
\label{actiogas}
\eeq
where $x = \alpha \ln u$, $y=\beta \ln v$ and $\dot x^2 + \dot y^2 =1$, as indicated earlier, because 
the metric is flat. This is a consequence of the coordinate invariance of the formalism.

\subsection{The Van der Waals gas}

The treatment of the Van der Waals gas in similar, although the evolution equations appear 
more complicated at first sight. In the representation above, the Van der Waals gas is defined 
by the equations of state
\bea
\left(u+\frac av\right)p_u = \frac{gk}2\cr\cr
\left(p_v + \frac{ap_u}{v^2} \right) (v-b) = k
\label{statevdw}
\eea
so \eqref{geng} turns into
\beq
g^{\mu\nu} = \left(\begin{matrix}
\frac 1{\alpha^2}\left(u+\frac{a}{v}\right)^2 + \frac{a^2(v-b)^2}{\beta^2 v^4} & \frac{a(v-b)^2}
{\beta^2 v^2}\cr\cr
\frac{a(v-b)^2}{\beta^2v^2} & \frac{(v-b)^2}{\beta^2}
\end{matrix}\right)
\label{gvdw}
\eeq
and on the physical hypersurface, $F=0$,
\beq
\lambda = - \sqrt{\frac{g^2k^2}{4\alpha^2}+\frac{k^2}{\beta^2}}.
\label{lvdw}
\eeq
With this we find the following evolution equations 
\bea
&& \dot u = -\frac 1\lambda \left[\frac{p_u}{\alpha^2}\left(u+\frac av\right)^2 + \frac a{\beta^2v^2}
\left(p_v + \frac{ap_u}{v^2}\right)(v-b)^2\right]\cr\cr
&&\dot v = -\frac 1{\beta^2\lambda} \left(p_v + \frac{ap_u}{v^2}\right)(v-b)^2\cr\cr
&&\dot p_u = \frac 1{\alpha^2\lambda} p_u^2\left(u + \frac av\right)\cr\cr
&&\dot p_v = \frac 1\lambda\left[-\frac{ap_u^2}{\alpha^2v^2}\left(u+\frac av\right) -\frac{2ap_u}
{\beta^2v^3}\left(p_v + \frac{ap_u}{v^2}\right)(v-b)^2+ \frac 1{\beta^2}\left(p_v + \frac{ap_u}{v^2}
\right)^2(v-b)\right]
\eea
Now, employing the middle two equations, the first and last can be put in the form
\bea
&&\frac d{d\tau}\left(u+\frac av\right) = - \frac{p_u}{\alpha^2\lambda} \left(u+\frac av\right)^2\cr\cr
&&\frac d{d\tau}\left(p_v + \frac{ap_u}{v^2}\right) = \frac 1{\beta^2\lambda} \left(p_v + \frac{ap_u}
{v^2}\right)^2(v-b)
\eea
from which it is clear that \eqref{statevdw} is recovered and, furthermore, the solutions
\bea
u &=& - \frac av + \left(u_0+\frac a{v_0}\right) \exp\left[\frac{gks}{2\alpha^2\lambda^2}\right]\cr\cr
v &=& b + (v_0-b)\exp\left[\frac{ks}{\beta^2\lambda^2}\right],
\label{solsvdw}
\eea
which may be inverted to obtain
\beq
s = k \ln \left[ \left(\frac{u+\frac av}{u_0+\frac a{v_0}}\right)^{g/2} \left(\frac{v-b}{v_0-b}\right)\right]
\eeq
once use is made of \eqref{lvdw}. As before, \eqref{solsvdw} can be recovered as geodesics of the metric \eqref{gvdw}, or from 
the action in \eqref{actiogas} with 
\beq
x=\alpha \ln \left(u + \frac av\right),~~ y = \beta \ln (v-b).
\eeq

\subsection{Paramagnetism}

For definiteness consider spin $\frac 12$ paramagnetism for which the first law will read
\beq
ds = \frac 1T dh - \frac bT dm
\eeq
where $h=u+mb$ is the specific enthalpy, $m$ is the magnetization and $b$ is the magnitude of an 
applied, external magnetic field. In this representation the configuration space is made of the 
pair $(h,m)$ and the 
conjugate forces are respectively $p_h = 1/T$ and $p_m = -b/T$. A paramagnetic material does not have
interactions between its dipoles and its enthalpy is completely independent of the magnetization, 
depending only on $T$. The simplest model of this non magnetic contribution would be to imagine that the 
paramagnetic molecules are oscillators, oscillating about their equilibrium positions. Therefore, if 
we suppose the enthalpy to be given by the law of equipartition, the material is completely 
characterized by the relations
\bea
m &=& \mu \tanh\frac{\mu b}{kT}\cr\cr
hp_h &=& gk
\eea
where $\mu$ is the magnetic moment of the particles. We can write these relations in terms of the 
phase space variables in the form
\beq
p_m = -\frac k\mu\tanh^{-1} \left(\frac m\mu\right),~~ hp_h = gk
\label{condpm}
\eeq
They suggest the family of metrics
\beq
g^{\mu\nu} = \left(\begin{matrix}
\frac{h^2}{\alpha^2} & 0\cr
0 & \frac{1}{\beta^2}[\atanh(m/\mu)]^{-2}
\end{matrix}\right)
\label{gpm}
\eeq
for which 
\beq
\lambda = -\sqrt{\frac{g^2k^2}{\alpha^2} + \frac{k^2}{\mu^2\beta^2}}
\label{lpm}
\eeq
on the physical hypersurface, $F=0$. We find the equations
\bea
&&\dot h = -\frac{h^2p_h}{\alpha^2\lambda}\cr\cr
&&\dot m = -\frac{p_m}{\beta^2\lambda[\atanh(m/\mu)]^2} \cr\cr
&&\dot p_h = \frac{hp_h^2}{\alpha^2\lambda}\cr\cr
&&\dot p_m = -\frac{p_m^2}{\beta^2\lambda\mu(1-m^2/\mu^2)[\atanh(m/\mu)]^3}
\eea
from which follow \eqref{condpm} and the solutions
\bea
&&h = h_0 \exp\left[-\frac{gks}{\alpha^2\lambda^2}\right],\cr\cr 
&&\frac m{2\mu} \ln \left[\frac{1+m/\mu}{1-m/\mu}\right] + \frac 12 \ln \left[1-\frac{m^2}{\mu^2}\right] 
= - \frac{ks}{\beta^2\mu^2\lambda^2} + s_0.
\eea
The second determines the magnetic contribution to the specific entropy. That the entropy 
is the sum of the two contributions follows by simply eliminating the arbitrary constants, $\lambda$, 
$\alpha$ and $\beta$ in the two equations above, using \eqref{lpm}. Once again, these solutions are geodesics 
of \eqref{gpm}. They can also be obtained from the action in \eqref{actiogas} with
\beq
x = \alpha \ln h,~~ y = \beta \left\{m \tanh^{-1} (m/\mu) + \frac \mu 2 \ln[1-(m/\mu)^2]\right\}.
\eeq

\subsection{Kerr Black Hole}

The Kerr black solution is one of four black hole solutions in general relativity and describes a 
neutral, rotating black hole. It is completely characterized by its axisymmetry and by two parameters, 
{\it viz.,} the mass of the black hole, $M$, and its angular momentum, $J$. It is described 
by the spacetime metric (we take $c=1$ and $G=1$)
\beq
ds^2 = \left(1-\frac{r_s r}{\rho^2}\right) dt^2 + \frac{2r_s r a\sin^2\theta}{\rho^2} dtd\varphi
- \frac{\rho^2}\Delta dr^2 -\rho^2 d\theta^2 - \left(r^2 +a^2 + \frac{r_s r a^2}{\rho^2} \sin^2\theta
\right) d\varphi^2
\eeq
where $r_s$ is the Schwarzschild radius, $r_s = 2M$, and
\bea
&&a = J/M\cr
&&\rho^2 = r^2 + a^2 \cos^2\theta\cr
&&\Delta = r^2 - r_s r + a^2
\eea
There are two surfaces of interest: an inner surface (the event horizon) occurring at 
\beq
r_h = \frac 12 (r_s + \sqrt{r_s^2-4a^2})
\eeq
and an outer surface of infinite redshift at
\beq
r_e = \frac 12 (r_s + \sqrt{r_s^2-4a^2\cos^2\theta})
\eeq
The two horizons meet at $\theta=0$ and the region between them is called the ergosphere. Within
the ergosphere, a test particle must co-rotate with the mass $M$, with the angular velocity
\beq
\Omega = \frac a{r_h^2+a^2}.
\label{avkbh}
\eeq
A quantum field placed in this background is well known to acquire a temperature. This temperature is 
proportional to the acceleration of the null Killing vector on the horizon, known as the surface 
gravity, $\kappa$, of the hole,
\beq
T = \frac{\kappa}{2\pi} = \frac{r_h^2-a^2}{4\pi r_h(r_h^2+a^2)}.
\label{Tkbh}
\eeq
The first law of black hole thermodynamics can be written in the form
\beq
dS = \frac 1T dM - \frac \Omega T dJ .
\eeq
From the thermodynamic point of view, the configuration space is two dimensional, spanned by 
the Arnowitt-Deser-Misner mass, which plays the role of the internal energy, and the angular 
momentum. The corresponding conjugate forces are $p_M = 1/T$ and $p_J= -\Omega/T$. Although these 
are natural variables for the system, it is convenient to transform to the configuration space 
$(r_h, a)$, writing the first law of black hole thermodynamics in terms of these variables as 
\beq
ds = p_h dr_h + p_a da
\eeq
instead. The equations of state, \eqref{avkbh} and \eqref{Tkbh}, are then equivalent to the 
statements that $p_h = 2\pi r_h$ and $p_a = 2\pi a$ and one finds
\beq
g^{\mu\nu} = \left(\begin{matrix}
\frac{1}{\alpha^2 r_h^2} & 0 \cr
0 & \frac{1}{\beta^2a^2}
\end{matrix}\right),
\label{gkbh}
\eeq
which yields 
\beq
\lambda = -2\pi \sqrt{\frac 1{\alpha^2}+\frac 1{\beta^2}}.
\label{lkbh}
\eeq
on the physical hypersurface. The evolution equations 
\bea
&&\dot r_h = - \frac{p_h}{\alpha^2 \lambda r_h^2}\cr\cr
&&\dot a = - \frac{p_a}{\beta^2\lambda a^2}\cr\cr
&&\dot p_h = -\frac{p_h^2}{\alpha^2\lambda r_h^3}\cr\cr
&&\dot p_a = -\frac{p_a^2}{\beta^2\lambda a^3}
\eea
have the solutions
\beq
\frac{p_h}{r_h} = \text{const.},~~ \frac{p_a}{a} = \text{const.},~~ r_h^2 = \frac{4\pi s}{\alpha^2
\lambda^2}. + r_{h0}^2,~~ a^2 =  \frac{4\pi s}{\beta^2\lambda^2} + a_0^2
\eeq
and the black hole entropy
\beq
s = \pi (r_h^2+a^2) + s_0
\eeq
is recovered as the simple sum of contributions from $r_h^2$ and $a^2$ by eliminating the constants 
using \eqref{lkbh}. As in our previous examples, these solutions are geodesics of \eqref{gkbh} and 
can also be obtained from the action in \eqref{actiogas} with
\beq
x = \alpha^2 r_h^2,~~ y = \beta^2 a^2
\eeq

\section{Discussion}

If spacetime is indeed emergent, its microscopic degrees of freedom quite possibly live at scales on 
the order of the Planck length. Therefore, not only do we presently have no experimental access to to 
them but we are most likely never to have it. It would seem that the best we could hope to empirically 
justify is an ever more precise description of the thermodynamics of this microstructure and, in turn, 
such a sharpened description may eventually lead to a better understanding of the fundamental 
constituents of spacetime. Progress along these lines can be made only once we have a clearer picture 
of the connections between our current geometric description of spacetime and thermodynamics. In this 
paper we have begun to address this issue by taking a closer look at thermodynamics. 

We were able to show that at least a subset of the possible transformations on the thermodynamic phase 
space lead to the description of thermodynamics processes as geodesics of a family of metrics defined 
by the equations of state of the substance. In the entropy representation, it is the entropy that 
serves as thermodynamic ``time''. In general, it is the ``preferred'' function $f$ that plays this 
role. Although all the metrics in the examples we have considered are flat there is no reason 
to expect that this is universally so, particularly in higher dimensional systems, eg., systems 
with variable contents or charged and rotating black holes. It would be of considerable interest 
to study the non-trivial geometry of these systems and correlate their geometric properties with 
their thermodynamic behavior. Such work, based on the Ruppeiner or Weinhold metric, has been 
attempted \cite{rup08,rup10,bel10} and our work can be seen as providing additional motivation for
it.

It is interesting that the family of relevant metrics is completely recovered simply from the equations 
of state. This is in fact what one should expect in an emergent picture of spacetime: that the 
geometric formulation is simply a way of specifying the spacetime equations of state. We hope that, 
with more work, this insight may help to ``design'' geometric models of emergent gravity that are 
better behaved in the ultraviolet and perhaps even to eventually understand spacetime's microstructure 
better in an information theoretic way.

\end{document}